\documentclass[conference]{IEEEtran}
\IEEEoverridecommandlockouts

\usepackage{derivative}
\usepackage{cite}
\usepackage{bbm}
\usepackage{amsmath,amssymb,amsfonts}
\usepackage{algorithmic}
\usepackage{graphicx}
\usepackage{textcomp}
\usepackage{xcolor}
\usepackage{tabularx}
\usepackage{xcolor}
\usepackage{balance}
\definecolor{links}{rgb}{0.3,0,0}   
\definecolor{urls}{rgb}{0,0,0.8}    
\definecolor{cites}{rgb}{0,0,0.6}   
\usepackage[colorlinks,hyperindex,linkcolor=links,citecolor=cites,urlcolor=urls]{hyperref} 

\definecolor{silver}{cmyk}{0,0,0,0.3}
\definecolor{navy}{cmyk}{0.8,0.5,0,0}
\definecolor{lightblue}{cmyk}{0.35,0.11,0,0}
\definecolor{orange}{cmyk}{0,0.57,0.86,0}
\definecolor{yellow}{cmyk}{0,0,0.9,0.0}
\definecolor{reddishyellow}{cmyk}{0,0.22,1.0,0.0}
\definecolor{lightred}{cmyk}{0,0.820,0.753,0.0}
\definecolor{black}{cmyk}{0,0,0.0,1.0}
\definecolor{white}{cmyk}{0,0,0.0,0}
\definecolor{purple}{cmyk}{0.64,0.83,0,0}
\definecolor{darkYellow}{cmyk}{0,0,1.0,0.5}
\definecolor{darkSilver}{cmyk}{0,0,0,0.1}
\definecolor{lightyellow}{cmyk}{0,0,0.3,0.0}
\definecolor{lighteryellow}{cmyk}{0,0,0.1,0.0}
\definecolor{lightestyellow}{cmyk}{0,0,0.05,0.0}
\definecolor{darkblue}{cmyk}{0.98,0.89,0,0.11}
\definecolor{bluel1}{cmyk}{0.5,0.05,0.05,0.05}
\definecolor{darkred}{cmyk}{0,0.89,0.7,0.55}
\definecolor{magenta}{rgb}{1.0, 0.0, 1.0}
\definecolor{cyan}{rgb}{0.0, 1.0, 1.0}

\usepackage{dsfont}
\usepackage{footnote}
\usepackage{float}
\usepackage{cite}
\usepackage{comment}

\usepackage{vmr-symbols-vecbold}
\usepackage{standard-macros}

\begin{document}
\safemath{\throughput}{899.9}
\title{RFNoC-Based FPGA Offloading for Fully Programmable PHY Acceleration \\
}

\author{\IEEEauthorblockN{A. Oguz Kislal\textsuperscript{1}, Osman Mert Yilmaz\textsuperscript{1,2}, Bengu Bilgic Keskin\textsuperscript{1}, Ibrahim Hokelek\textsuperscript{1},
Ali Gorcin\textsuperscript{1,2}}
\IEEEauthorblockA{\textsuperscript{1}\textit{Communications and Signal Processing Research (HİSAR) Laboratory, TÜBİTAK BİLGEM, Kocaeli, Turkiye}\\
\textsuperscript{2}\textit{Department of Electronics and Telecommunications Engineering Istanbul Technical University, Istanbul, Türkiye}\\
Emails: \{ahmet.kislal, mert.yilmaz, bengu.bilgic, ibrahim.hokelek\}@tubitak.gov.tr, ali.gorcin@gmail.com}
}

\maketitle

\begin{abstract}

Hardware acceleration has emerged as a key research topic for supporting computationally intensive signal processing and artificial intelligence applications in 6G research and development studies. This paper presents an RF Network-on-Chip (RFNoC)–based hardware acceleration framework that offloads key physical layer procedures to a field-programmable gate array (FPGA). The proposed design accelerates procedures, including low-density parity-check codes (LDPC) encoding and decoding, rate matching and unmatching, interleaving and deinterleaving, scrambling and descrambling, and log-likelihood ratio estimation. The accelerator is integrated directly into the OpenAirInterface radio access network software, enabling simultaneous use of the FPGA as driver of the radio front-end and a high-throughput accelerator. The proposed system is validated through real-time experiments with a commercial smartphone successfully connecting to the network. The implementation results demonstrate that a throughput of about 900 Mbps is achievable using a moderate FPGA resource utilization.
\end{abstract}

\section{Introduction}
Vehicular communication systems place stringent requirements on the underlying cellular infrastructure, particularly in terms of latency, reliability, and real-time processing under high mobility. In vehicle-to-infrastructure scenarios, rapid physical (PHY) layer decoding and timely feedback at the base station are essential to support safety-critical and latency-sensitive services. These requirements motivate efficient, low-latency PHY layer implementations at the roadside units and edge-deployed base stations. Furthermore, as future vehicular communication systems evolve toward 6G-enabled vehicle-to-everything (V2X), even tighter latency constraints and higher data rates are expected at the infrastructure side \cite{Mario2021}.

To meet these stringent requirements in practice, open-source radio access network (RAN) stacks such as OpenAirInterface (OAI)~\cite{Nikaein2014} have been widely adopted in research and prototyping \cite{Fuat2025,Sever2025}. A common approach for achieving a high-throughput is to run these stacks on general-purpose processors with advanced vector extensions 512-bit (AVX-512) accelerators \cite{Yi2019}. While this approach is effective, it typically relies on expensive multi-core processor hardware, and the resulting accelerator design lacks re-programmability. A viable alternative is to use a field-programmable gate array (FPGA), which provides a highly flexible and programmable hardware platform for reconfiguring functionality to meet application-specific requirements. Furthermore, FPGAs facilitate the development and prototyping of high-speed application-specific integrated circuits (ASICs) for 6G studies.

Software-defined radio (SDR), which has been preferred for next-generation wireless network studies, typically includes a reprogrammable processor and FPGA and RF front-end capabilities.  For universal software radio peripheral (USRP) X410, a Zynq UltraScale+ FPGA is included and mostly under-utilized, making them suitable for offloading some or all PHY layer operations. Moreover, USRP X410 comes with RF Network-on-Chip (RFNoC), an FPGA-based design framework that significantly improves the flexibility and performance of software-defined radio (SDR) systems. Instead of fixed processing chains, it enables developers to implement custom signal-processing modules directly on the FPGA while abstracting and simplifying interface designs, e.g. ADC/DAC, small form-factor pluggable (SFP+) etc.~\cite{Braun2016}.

In \cite{Wei2022}, the authors analyzed the OAI framework and show that the PHY layer processing cost for uplink traffic is notably higher than that of comparable downlink rates. In another study, a graphical processing unit (GPU) accelerated inline hardware design for both uplink and downlink is proposed~\cite{Kundu2024}. A key limitation of these works is that they mainly demonstrate the acceleration of individual functions, rather than providing a full evaluation of hardware resource requirements across different PHY layer blocks. 
Similarly, the LDPC and polar encoder/decoder are offloaded to the FPGA within a USRP through the RFNoC architecture in \cite{Tuerxun2024}. They report the latency results by integrating their channel coding accelerator with the OAI software. However, none of these works report the performance results while the end-to-end network is under operation.  Instead, their experiments rely on the PHY simulator included in the OAI emulator. As we shall discuss, using the RFNoC framework to stream high-throughput data between the FPGA and the RF System on Chip (RFSoC) while simultaneously performing hardware acceleration introduces significant challenges.


In this paper, an RFNoC–based hardware acceleration framework that offloads key PHY layer procedures to an FPGA is presented. The main contributions of this work are summarized as follows.
\begin{itemize}
    \item We design and develop an FPGA-based digital signal processing chain for 5G New Radio (NR) PHY layer procedures, synthesized on the Zynq UltraScale+ and operating at 250 MHz and 500 MHz clock frequency. The experiment results show that a throughput of about 900 Mbps is achievable.
    \item  The proposed accelerator is integrated into the OAI software through the RFNoC framework. We demonstrate that a commercial phone is connected to an 5G network emulated through OAI software while the accelerator executes key PHY layer tasks and drive the RF front-end simultaneously. We will release an open-source library that enables the RFNoC architecture to be integrated with the OAI software.
    \item The proposed design provides an end-to-end, flexible, and programmable platform in which the RFNoC-based FPGA accelerator performs the PHY-layer tasks, while the OAI emulator executes the higher-layer procedures. It can be extended to support emerging 6G use cases, such as integrated sensing and communication (ISAC) and cell-free MIMO.
\end{itemize}





\section{Hardware Accelerated PHY Layer Procedures}
\begin{figure*}
    \centering
    \includegraphics[width=2\columnwidth]{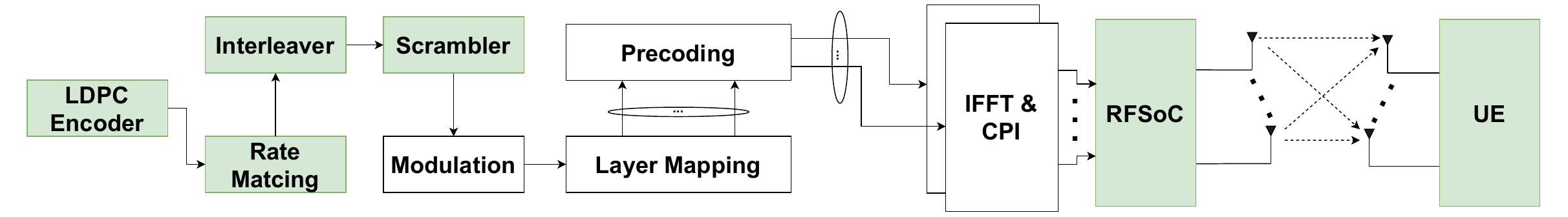}
    \caption{Block diagram of PHY layer encoder procedures. Here, green blocks are offloaded to FPGA.}
    \label{fig:Sec2_phyEnc}
\end{figure*}

\begin{figure*}
    \centering
    \includegraphics[width=1.5\columnwidth]{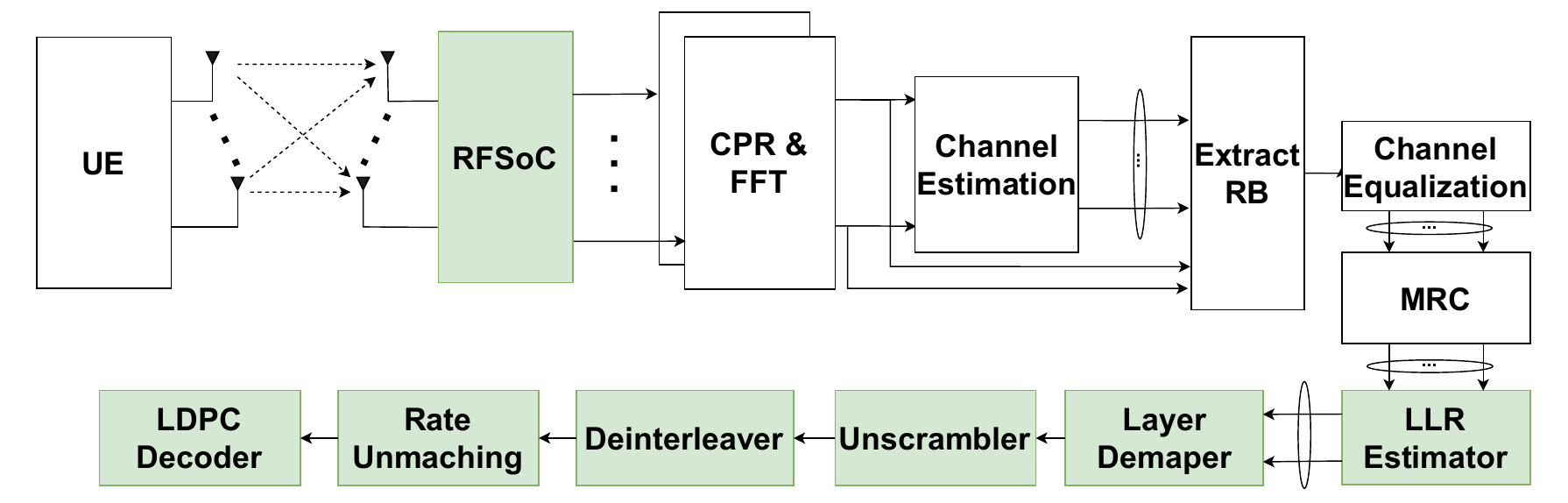}
    \caption{Block diagram of PHY layer decoder procedures. Here, green blocks are offloaded to FPGA.}
    \label{fig:Sec2_phyDec}
\end{figure*}

In this section, we describe the designed accelerated PHY layer procedures whose block diagrams are shown in Fig. \ref{fig:Sec2_phyEnc} and Fig. \ref{fig:Sec2_phyDec}, respectively \cite{Saaifan2021}.
Note that the encoding processes for 5G NR are defined in \cite{3gpp38_212}. However, for the sake of completeness, we briefly summarize each encoding step and its key parameters when presenting our design. The decoding procedures are left unspecified in the standard since the decoder is expected to be derived from the encoding process. In addition, hybrid automatic repeat request (HARQ) introduces a design trade-off between achievable throughput and the amount of memory that must be allocated by the decoder. 

In our design, we utilize 32-bit vectors for the encoding procedures and both 32-bit and 128-bit vectors for the decoding procedures. All buffers and memories, unless otherwise specified, are implemented using block RAMs (BRAMs). Since BRAMs support simultaneous read and write operations at different addresses within the same clock cycle, they play a crucial role in achieving high throughput.
All BRAM outputs are registered to reduce the logic-path delay. In some cases, we also registered the BRAM input data and address pointer to satisfy timing requirements along data paths. 

\subsection{Encoding Procedure}
\label{sec:EncodingProcedure}
The encoding procedures implemented on the FPGA include LDPC encoding, rate matching, interleaving, and scrambling. These stages are typically followed by modulation, which could also be executed on the FPGA. However, performing modulation on the FPGA creates a significant data-movement penalty: for a modulation order of $Q_m$, each group of $Q_m$ bits is expanded into a complex symbol represented by a 16-bit real and 16-bit imaginary component (32 bits total). Even at the highest modulation order defined in 5G NR $(256-QAM, Q_m=8)$, this results in a fourfold increase in the amount of data that must be transferred to the CPU. For this reason, we keep the modulation on the CPU side.

\subsubsection{LDPC Encoder}
LDPC coding in 5G NR is based on two predefined base graphs (BG), namely BG1 for large block sizes and BG2 for smaller ones. These BGs provide compact templates of the parity-check matrix. The actual matrix used for encoding is obtained by lifting the BG by a factor $Z_c$, where $Z_c$ defines the dimension of each submatrix that replaces an entry in the BG. For BG1, the resulting input blocklength is $K = 10Z_c$ and the codeword length is $N_{cb} = 50Z_c$. For BG2, the corresponding values are $K = 22Z_c$ and $N_{cb} = 66Z_c$.
If the transport-block segment length $K'$ is smaller than $K$ ($Z_c$ is selected so that $K \ge K'$), the input vector is zero-padded by $F = K - K'$ bits. The full codeword length of the underlying parity-check matrix is $N = N_{cb} + 2Z_c$ for\footnote{According to the 5G NR specification \cite{3gpp38_212}, the first $2Z_c$ bits of the encoded output are always punctured, as they are sufficiently protected by the remaining part of the codeword.} both BGs.

\subsubsection{Rate Matching}
The rate-matching procedure reduces the $N_{cb}$ encoded bits to $E_r$ bits for each block. To implement this, we first write the $N_{cb}$ bits into a circular buffer of length $N_{cb}$. From a predefined starting position, $E_r$ bits are then read sequentially from this buffer. When HARQ is activated, different portions of the codeword are selected simply by modifying the starting position.

Our implementation uses a single BRAM36, which can store up to $792$ 32-bit vectors, together with 16-bit barrel shifters to handle the bit shifts required when the circular read operation wraps around the buffer and when $N_{cb}$ is not a multiple of 32.


\subsubsection{Interleaver}
The interleaving stage reorganizes the rate-matched bits so that those mapped to the same modulation symbol are spread across the codeword, increasing robustness against burst errors. Starting from the $E_r$ input bits, the interleaver produces groups of $Q_m$ bits. Each $Q_m$-bit group is constructed by selecting the $\{k + \ell(E_r/Q_m)\}$th bits, where $k \in \{1, \ldots, E_r/Q_m\}$ and $\ell \in \{0, \ldots, Q_m - 1\}$.

To implement this operation, we use a BRAM36 storing up to $1024$ 32-bit vectors, along with 16-bit barrel shifters to manage bit concatenation when $E_r/Q_m$ is not a multiple of 32. Note that rate matching does not always correspond to puncturing; in some configurations, portions of the codeword may be repeated rather than removed.

\subsubsection{Scrambler}
In 5G NR, scrambling is used to apply a cell- and transmission-specific pseudo-random sequence to the coded bits, preventing deterministic patterns from appearing before modulation. The scrambling sequence is a gold sequence whose initialization depends on the radio network temporary identifier (RNTI) and the cell ID. In our implementation, the gold sequence is generated in 32-bit segments, and each 32-bit segment is XOR-ed with the corresponding 32-bit portion of the rate-matched data on every clock cycle.

\subsection{Decoding Procedure}
\label{sec:decoding}
For decoding offlading, except for LLR estimation, all functions correspond directly to the encoding operations described in Section~\ref{sec:EncodingProcedure}. Therefore, in the following subsections, we concentrate primarily on their digital design and FPGA implementation. Here, we used 250 MHz clock rate for all the logic operations except LDPC decoding. For LDPC decoding, by utilizing the hardened LDPC decoder IP core in the FPGA, we used 500 MHz clock rate.

\subsubsection{LLR Estimator}
LLR estimation computes the soft bit metrics from the equalized constellation symbols, providing reliability information essential for LDPC decoding. It is given as
\begin{equation}
\label{llr}
\text{LLR}_b = \log \ltrp{ \frac{ \Prob
(b = 1 \given r) }{ \Prob(b = 0 \given r) } } 
\end{equation}
where $r$ is the received equalized symbol, and $b = \{0, 1\}$  is the binary variable. Since evaluating exact LLR requires $2^{Q_m}$ distance  evaluation over the constellation for $E_r$ bits, it is computationally expensive. To simplify LLR estimation, max-log approximation is commonly used in the literature \cite{coding_book}. Max-log approximated LLR estimation is given as
\begin{equation}
\text{LLR}_b \simeq 
\frac{1}{2\sigma^{2}}
\ltrp{
    \min_{s \in \mathcal{S}(b = 1)} \abs{r - s}^{2}
    -
    \min_{s \in \mathcal{S}(b = 0)} \abs{r - s}^{2}
}
\label{llr_approx}
\end{equation}
where $\sigma^2$ is the variance of noise and  $\mathcal{S}$ is the set of constellation symbols.
For our constellation, the in-phase and the quadrature components of quadrature amplitude modulation (QAM) symbols can be treated as two independent pulse amplitude modulation (PAM) symbols. Making use of the symmetric structure of gray-mapped QAM constellations, LLR calculations are simplified into piecewise-linear functions \cite{llr2}. In \cite{llr1}, authors further approximates these piecewise-linear functions as 
\begin{align}
\label{eq:LLR0}
\mathrm{LLR}_{b_0} &\simeq -r_I \frac{A_{Q_m}}{\sigma^2}  \\
\mathrm{LLR}_{b_1} &\simeq -r_Q  \frac{A_{Q_m}}{\sigma^2} \\
\mathrm{LLR}_{b_2} &\simeq \big(|r_I| - B_{Q_m}\big)  \frac{A_{Q_m}}{\sigma^2} \\
\mathrm{LLR}_{b_3} &\simeq \big(|r_Q| - B_{Q_m}\big)  \frac{A_{Q_m}}{\sigma^2} 
\label{eq:LLR3}
\end{align}
\begin{align}
\label{eq:LLR4}
\mathrm{LLR}_{b_4} &\simeq \big(\,\big||r_I| - B_{Q_m}\big| - C_{Q_m}\big)  \frac{A_{Q_m}}{\sigma^2} \\
\mathrm{LLR}_{b_5} &\simeq \big(\,\big||r_Q| - B_{Q_m}\big| - C_{Q_m}\big)  \frac{A_{Q_m}}{\sigma^2} \\
\mathrm{LLR}_{b_6} &\simeq \big(\,\big|\big||r_I| - B_{Q_m}\big| - C_{Q_m}\big| - D_{Q_m}\big)  \frac{A_{Q_m}}{\sigma^2}  \\
\mathrm{LLR}_{b_7} &\simeq \big(\,\big|\big||r_Q| - B_{Q_m}\big| - C_{Q_m}\big| - D_{Q_m}\big) \frac{A_{Q_m}}{\sigma^2}
\label{eq:LLR7}
\end{align}
where  $r_I = \Re(r)$, $r_Q = \Im(r)$ are the in-phase and quadrature part of the received signal, $b_k$, $k \in \{0,\ldots,Q_m\}$ is the $k$th bit of a symbol, and $A_{Q_m}$, $B_{Q_m}$, $C_{Q_m}$ and $D_{Q_m}$ are constants defined in Table~\ref{tab:LLR_params}.
\begin{table} \centering \caption{LLR approximation parameters}
\begin{tabular}{c c c c c c}
$Q_m$ & Modulation & $A_{Q_m}$ & $B_{Q_m}$ & $C_{Q_m}$ & $D_{Q_m}$ \\ [0.5ex]  
\hline   
2 & QPSK & $\dfrac{2}{\sqrt{2}}$ & -- & -- & -- \\ [2ex]   
4 & 16QAM & $\dfrac{2}{\sqrt{10}}$ & $A_{Q_m}$ & -- & -- \\ [2ex]  
6 & 64QAM & $\dfrac{2}{\sqrt{42}}$ & $2A_{Q_m}$ & $A_{Q_m}$ & -- \\ [2ex]  
8 & 256QAM & $\dfrac{2}{\sqrt{170}}$ & $4A_{Q_m}$ & $2A_{Q_m}$ & $A_{Q_m}$
\end{tabular} \label{tab:LLR_params} \end{table}

In our implementation, each partial function in \eqref{eq:LLR0}–\eqref{eq:LLR7} is implemented using six 16-bit signed adders and eight multipliers mapped to eight DSP48E2 slices, arranged in a 13-stage pipeline. After every pipeline stage, the intermediate value is saturated to $16$ bits when required. At the final stage, the output is saturated to $6$ bits and sign-extended to $8$ bits. We represent LLRs using a symmetric fixed-point format with $4$ integer bits and $2$ fractional bits, resulting in a dynamic range of $[-7.75, 7.75]$. The final output of the module is organized into 32-bit vectors, with each LLR sign-extended to 8 bits and each vector holds 4 LLRs. To achieve a high throughput, we process two complex symbols per clock cycle.  
\subsubsection{Descrambler}
The descrambler follows almost the same architecture as the scrambler. The main difference is that, while the scrambler processes 32 raw bits at its input, the descrambler operates on LLRs, where each 32-bit word contains four LLR values. Our descrambler can processes up to 16 LLRs per clock cycle. 
\subsubsection{Deinterleaver}
For deinterleaver, we aim to eliminate edge cases and achieve high throughput, at the cost of additional memory usage. To do so, we utilize four sets of $8$ buffers, where each buffer stores four 8-bit LLRs. In total, this structure requires 24 BRAM36.
Each of the $Q_m$ buffers stores incoming LLRs in the order needed to reverse the interleaving pattern. Using four parallel sets of $Q_m$ buffers allows us to read from all sets simultaneously and concatenate four 32-bit deinterleaved outputs into a single 128-bit vector.
Because the Xilinx LDPC Decoder IP core can process up to 16 LLRs (i.e., 128 bits) in parallel for each codeword, this design enables us to feed the decoder at its maximum throughput.

\subsubsection{HARQ Compliant Rate Unmatching}
Rate unmatching reverses the rate-matching process by placing the received LLRs back into their original positions in a circular buffer of length $N_{cb}$, as determined by the redundancy version $rv$. Once the LLRs are written to the buffer, the $F$ zero-padded bits are restored by assigning the minimum LLR value $-7.75$, while the punctured bits are reconstructed by setting their LLRs to $0$. HARQ is a critical feature to ensure error-free transmission of every packet by changing the coderate through re-transmissions. After every erroneous transmission of a packet, the receiver requests the re-transmission of the same packet with a different $rv$. As a result, in every consecutive round, a different part of the codeword is sent. The rate unmatching procedure needs to combine the received part of the codeword with the previously transmitted parts, essentially decreasing the coderate. 
Multiple packets are scheduled with different IDs, to increase the throughput of the system. As a result, after failing to decode a packet, the next available packets may have a different ID. 
To tackle with this, we used a memory implemented over 6 ultra RAM (URAM) that can store $16 \times 1652$ 128-bit vectors. We virtually divide it to $16$ circular buffer with length $N_{cb}$. To do so, the CPU first checks whether the received packet is a re-transmission or not. If it is a new packet, a free circular buffer is assigned to the codeword; otherwise, the previously allocated circular buffer is selected.
\subsubsection{LDPC Decoder}
The LDPC decoder iteratively reconstructs the transmitted codeword by exchanging reliability information between variable and check nodes. The LDPC decoder operates through iterative message-passing, most commonly using the low-complexity variants of sum-product algorithm (SPA) such as offset min-sum algortihm, which is particularly advantageous since it corrects the min-sum algorithm’s tendency to overestimate message magnitudes through a fixed offset. In our design, we used Xilinx’s LDPC decoder hardened IP core to implement the min-sum algorithm with an offset of $0.5$, the maximum of eight iterations, and allowing early termination either when the parity check is satisfied or when the hard decisions of the nodes remain unchanged.

\section{Hardware Acceleration Integration with OAI and RFNoC}
\begin{figure}
    \centering
    \includegraphics[width=0.95\columnwidth]{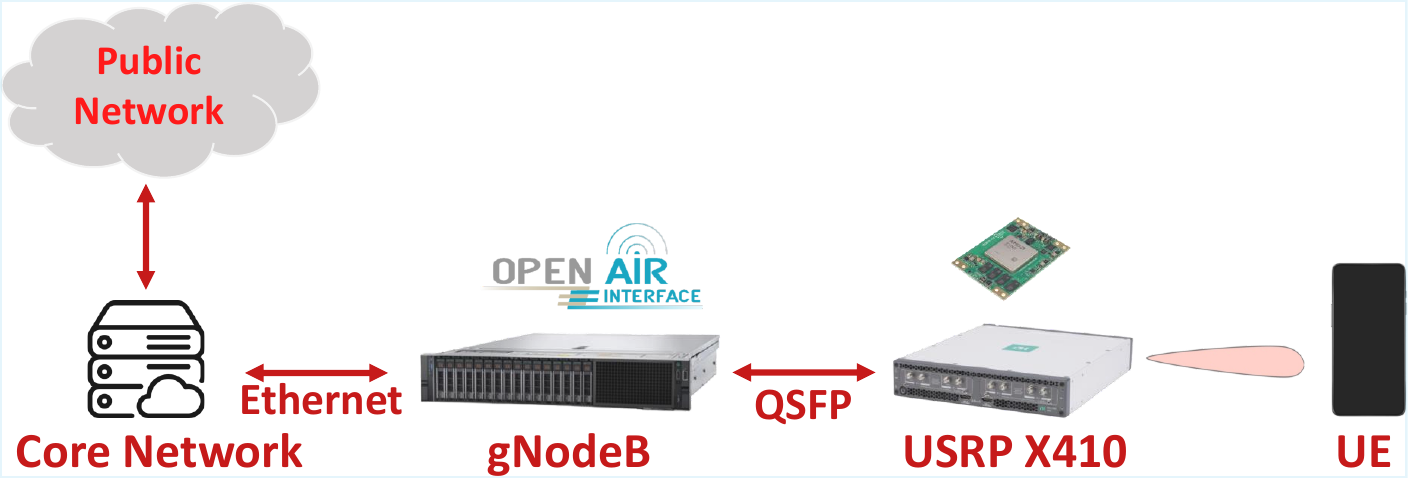}
    \caption{Testbed system model.}
    \label{fig:Sec2_fig1}
\end{figure}

Our setup comprises four main components: a core network, a gNodeb (gNB), a USRP X410 serving as both the radio front-end and a hardware accelerator, and a commercial smartphone acting as the user equipment (UE). The testbed system model is given in Fig. \ref{fig:Sec2_fig1}.
Here, the open-source OAI software, which is run over a workstation, is used to emulate both the core network and the gNB. 

\begin{figure}
    \centering
    \includegraphics[width=0.95\columnwidth]{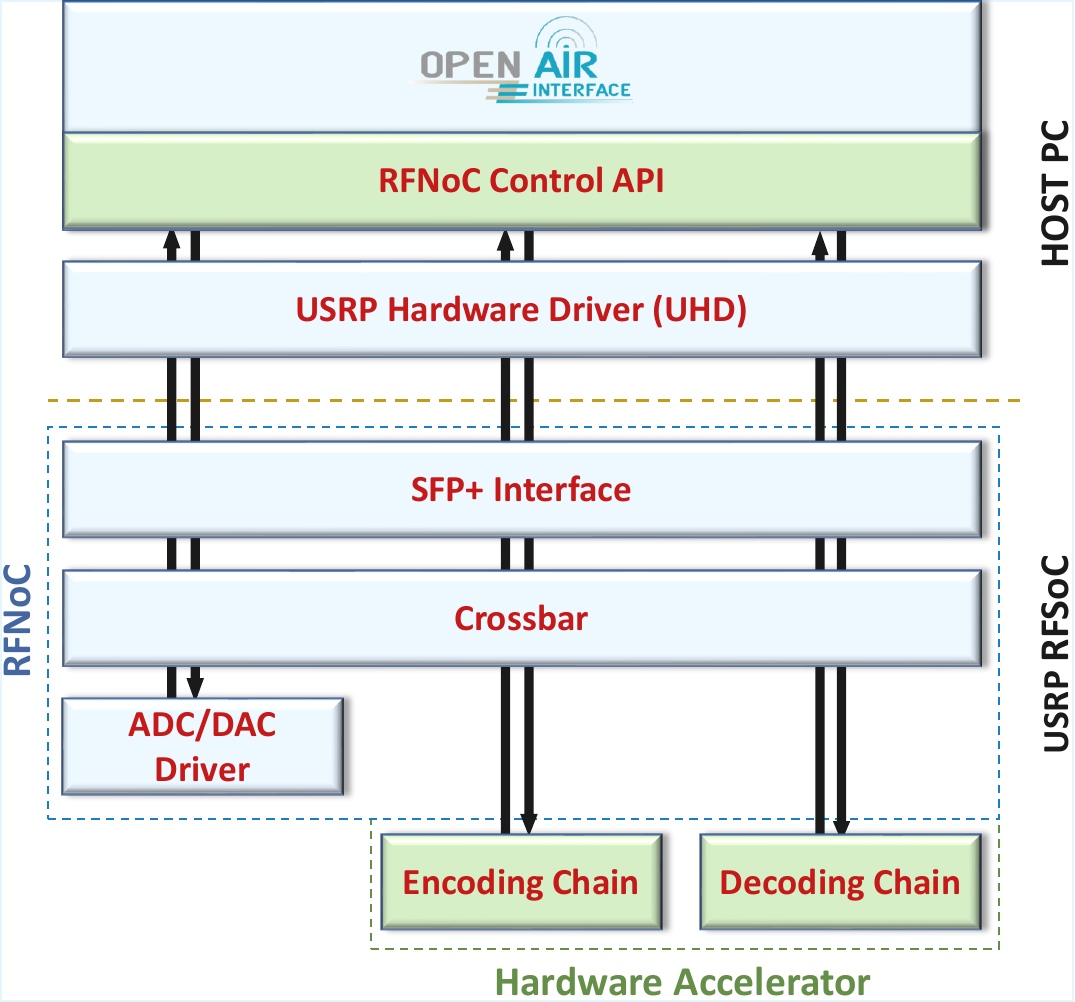}
    \caption{Functional block diagram of the hardware accelarated gNB.}
    \label{fig:Sec2_fig2}
\end{figure}


The functional block diagram of the proposed hardware accelerated gNB is given in Fig. \ref{fig:Sec2_fig2}. The host PC running the OAI software is connected to the RFSoC inside the USRP through SFP+ interface which connects to RFNoC. RFNoC is an FPGA-based framework that enhances the performance and flexibility of SDR systems by allowing custom signal processing blocks to be interconnected through a high-bandwidth on-chip network~\cite{Ettus2025}.
The \textit{multi-usrp} library, included in the RFNoC framework, provides an easy-to-use interface by abstracting the RFSoC driver from the underlying RFNoC architecture. Tools such as OAI and other 5G RAN software stacks rely heavily on this basic interface. However, this abstraction layer prevents access to user-defined signal-processing modules implemented within RFNoC.
To enable simultaneous use of the RFSoC platform and FPGA-based hardware acceleration, we developed a new RFNoC control application-programming interface (API)\footnote{Our API library, along with its integration into OAI, is available at \url{https://github.com/OsmanMertYilmaz/Hardware_Accelerated_OAI}.}, inspired by the interaction of the UHD driver with the OAI software. While our implementation has been tested primarily within the OAI, the API can be readily integrated into other emulation frameworks as well. 

\section{Experiment Results}
In this section, we present the implementation results of our setup consisting of a single USRP-X410 that has a Zynq UltraScale+ ZU28DR RFSoC and a linux workstation. The FPGA resource utilization is reported in Table \ref{table:FPGA_ResourceAlloc}.
These results indicate that more processing stages can be accommodated within the FPGA. Implementing the complete 5G NR PHY layer chain removes data offloading overhead and simplifies the overall system architecture.

\begin{table}[t] \centering \caption{FPGA Resource Utilization Summary}
\begin{tabular}{c | c c c}
Resource Type & Available & Used & Utilization \\ [0.5ex]  
LUT & $425280$ & $219137$ & $\%51.53$  \\ [0.2ex]
BRAM & $1080$ & $235.5$ & $\%23.47$\\ [0.2ex]
URAM & $80$ & $64$ & $\%80.00$ \\ [0.2ex]
DSP & $4272$ & $2042$ & $\%47.80$
\end{tabular} \label{table:FPGA_ResourceAlloc} \end{table}

\begin{table}[t] \centering \caption{Comparison of 5G NR Hardware Accelerators}
\begin{tabular}{c | c c c c c}
\textbf{Platform} & \textbf{Codeblock}  & \textbf{Throughput } & \textbf{Process }\\ [0.5ex] 
\hline \vspace{-1.2ex}\\ 
Our Setup & 20   & \throughput Mbps & See Sec. \ref{sec:decoding} \\ 
Our Setup & 40   & 900.1 Mbps & See Sec. \ref{sec:decoding} \\ 
RTX GPU \cite{Tarver2021} & 40 & $3964$ Mbps & LDPC Decoder \\
Xeon Gold 18 Core \cite{Yi2019}  & N/A & $1200$ Mbps & LDPC Decoder \\ 
Xeon Gold 1 Core \cite{Yi2019}  & N/A & $63.38$ Mbps & LDPC Decoder \\ 
Xilinx VC707 \cite{Papatheofanous2021} & 20 & $347$ Mbps & LDPC Decoder \\ 
\end{tabular} \label{table:ThroughputResults} \end{table}

Hardware acceleration over PHY layer is most commonly applied to LDPC decoding. As a result, directly comparing our work, which implements multiple decoding processes including the LDPC decoder, to existing results in the literature is not straightforward.
In our system, the initial latency, apart from LDPC decoding, is affected by the deinterleaving process. Throughput might also be affected, since the LDPC decoder can be continuously fed only when decoding spans multiple iterations\footnote{This limitation could be alleviated by utilizing multiple deinterleavers; however, having multiple blocks decoded in single-iteration successively is unlikely in practice.}.
Furthermore, while introducing a new hardware into the system (e.g., a GPU) increases the system complexity and power consumption, leveraging the existing resources available on RFSoC introduces no additional cost.
Nevertheless, to place our results in context, we report the throughput of state-of-the-art 5G LDPC decoding implementations alongside the maximum throughput achievable with our decoding chain, including all procedures explained in Sec. \ref{sec:decoding}, for the symbols generated at an SNR $10$~dB, $K=8448$, $N_{cb}=26112$, and $G=12672$ in Table~\ref{table:ThroughputResults}. In the literature, the highest reported throughput, $3964$~Mbps, is achieved using a GPU-based implementation \cite{Tarver2021}, followed by $1200$~Mbps from an 18-core CPU implementation using AVX-512 \cite{Yi2019}. Both results, however, are obtained under high-latency configurations that rely on data buffering. This largely offsets the impact of data-offloading. Under similar conditions, our system achieves a throughput of \throughput~Mbps. Note that this is a realistic benchmark for our system, as when the entire decoding chain is implemented on the FPGA, data offloading is not required. In this case, the received symbols can be fed directly into the decoding chain. 

\begin{figure}[t]
    \centering
    \includegraphics[width=0.85\columnwidth]{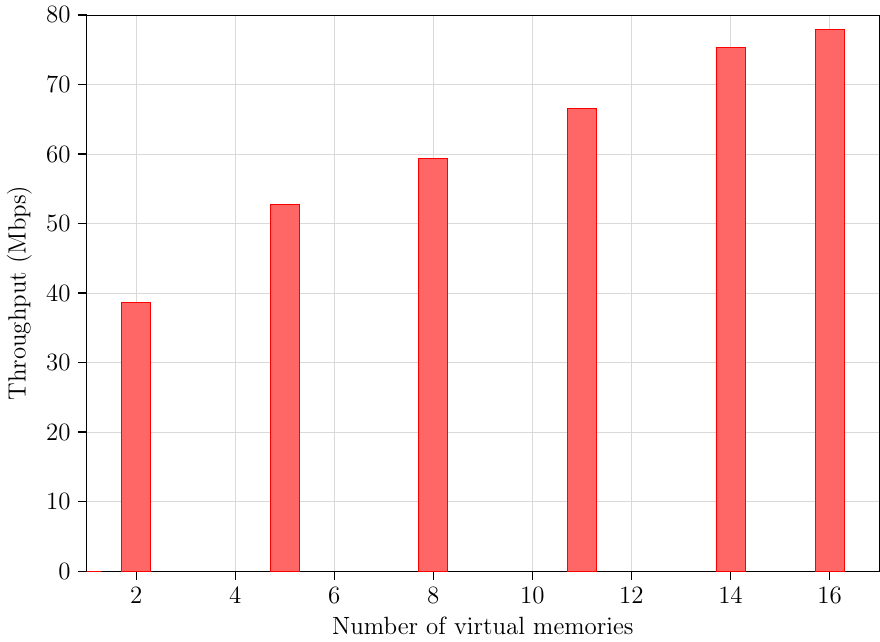}
    \caption{The average throughput as a function of number of virtual memories.} 
    \label{fig:Sec4_Result2}
\end{figure}

In Fig.~\ref{fig:Sec4_Result2}, we present the average throughput as a function of the number of virtual memories used in HARQ-compliant rate unmatching. The experiment is conducted using a hardware-accelerated SISO gNB emulator connected to a commercial smartphone, with a wall obstructing the line of sight between the gNB and the UE.
We first observe that when only a single virtual memory is used—effectively disabling HARQ and decoding each received block independently—a stable connection cannot be established. As expected, increasing the number of virtual memories leads to a significant improvement in throughput, reaching 77.9~Mbps when 16 virtual memories are used. Note that the throughput here is naturally limited by channel conditions and the number of transmission layers. In a single-antenna link, channel conditions are typically poor, and only one layer can be used for data transmission.


\section{Conclusion and Future Works}
We presented an end-to-end programmable platform based on an RFNoC-enabled hardware acceleration framework integrated with an open-source emulator. By selectively offloading key physical layer procedures to the FPGA, the proposed design enables the simultaneous use of the FPGA as both an RF transceiver and a high-throughput processing accelerator. Experimental results demonstrate that the proposed hardware accelerator achieves a throughput of approximately 900 Mbps.
Beyond performance gains, the proposed architecture provides a modular and extensible platform that facilitates rapid prototyping and evaluation of advanced PHY-layer techniques. The use of an RFNoC-based design allows new signal processing blocks to be seamlessly integrated and reconfigured, making the platform well suited for research on evolving cellular standards.

In future work, we plan to extend hardware offloading to the remaining physical layer procedures in order to further reduce end-to-end latency and improve overall system throughput. Additionally, the proposed framework will be expanded to support emerging 6G use cases, including integrated sensing and communication and cell-free MIMO, enabling experimental validation of next-generation wireless concepts on a flexible and reprogrammable hardware platform.

\end{document}